\documentclass[preprint,3p,12pt]{elsarticle}
\usepackage{amsfonts}
\usepackage{graphicx}
\usepackage{amssymb}
\usepackage{amsmath}
\usepackage{color} 
\usepackage{epsfig}

\biboptions{square,comma,sort&compress}

\journal{Annals of Physics}

\begin{document}

\begin{frontmatter}

\title{On Josephson effects in insulating spin systems}

\author[label1]{A.~Schilling}
\ead{aschilling@pysik.uzh.ch}
\author[label1]{H.~Grundmann}

\address[label1]{Physics Institute, University of Z\"urich, 
Winterthurerstrasse 190, CH-8057 Z\"urich}

\begin{abstract}

\noindent We discuss an experiment in which two magnetic 
insulators 
    that both show a field-induced magnetic-ordering transition are 
    weakly coupled to one another and are placed into an external 
    magnetic field. If the respective magnetic states can be 
interpreted 
    as phase coherent Bose-Einstein condensates of magnetic bosonic 
    quasiparticles, one expects the occurrence of Josephson effects. 
For two identical systems, the resulting d.c. Josephson effect 
formally represents a constant quasiparticle Josephson current across the weak link, 
which turns out to be unobservable in an experiment. For 
    magnetic insulators with different critical fields, a spontaneous 
alternating 
    quasiparticle current develops with a leading oscillation 
frequency 
    \(\omega_\text{a.c.}\) that is determined by the difference 
between 
    the critical fields. As a result of the coupling, additional 
    sidebands appear in the energy spectrum of the coupled device 
that 
    would be absent without phase coherence. We discuss the primary 
    conditions for such an effect to take place and conclude that its 
    detection can be feasible for a proper choice 
    of compounds with suitable 
    and realistic material parameters.

\end{abstract}

\begin{keyword}
    Tunneling, Josephson effect \sep Quantized spin models \sep 
Macroscopic quantum phenomena in magnetic systems \sep Josephson 
devices

\PACS 03.75.Lm \sep 75.10.Jm \sep 75.45.+j \sep 85.25.Cp

\end{keyword}

\end{frontmatter}

\section{Introduction}

Quantum spin systems in solids have been a subject of intense 
research, both theoretically and experimentally. A number of such 
systems show magnetic-field induced phase transitions at zero 
temperature that have been interpreted as a Bose-Einstein 
condensation of magnetic bosonic quasiparticles 
\cite{matsubara,batyev,affleck_antiferromag_bose,giamarchi_ladders,nikuni_tlcucl3_magnetization,ruegg_tlcucl3_nature,giamarchi_magneticisolator-bec,matsumoto_tlcucl3_pressure}.

In insulating dimerized spin-\(1/2\) systems, for example, 
one expects a condensation of triplet bosonic quasiparticles 
("triplons") above a certain critical field \(H_c\) 
\cite{giamarchi_ladders,nikuni_tlcucl3_magnetization} where the 
energy difference between the ground-state singlet and the lowest 
excited triplet states vanishes due to the Zeeman splitting.  In the 
language of magnetism, this condensation corresponds to a 
field-induced antiferromagnetic ordering that is associated with the 
appearance of a staggered magnetization \cite{matsubara,affleck_antiferromag_bose,giamarchi_ladders}. 
However, 
the stability of such a condensate 
\cite{dellamore_tlcucl3-symmetry-breaking,rakhimov_bec_instability} 
and even the applicability of the BEC concept 
\cite{kalita_dimer_magnet,bunkov_spin_superfluidity_magnonbec,mills_comment_magnon_bec} 
have been questioned.

Already known Bose-Einstein condensates (BEC) include superfluid 
helium \cite{allen_4he-flow,kapitza_4he-viscosity}, dilute atomic gas 
clouds \cite{anderson_atomic-bec} and pumped exciton-polariton and 
magnon condensates 
\cite{demokritov_magnon-bec,demidov_magnon-bec,deng_exciton-bec,kasprazak_polariton-bec}. 
Several experimental proofs for the existence of a macroscopic 
quantum state have been reported, e.g., from interference experiments 
\cite{hskinson_4he-interferometer,simmonds_3he-interference,andrews_gas-bec-interference}, from the observation of vortices 
\cite{hall_he-rotation,matthews_bec-vortex,lagoudakis_polariton_bec-vortex} and Josephson effects 
\cite{sukhatme_he-josephson,borovik_spin-josephson,pereverzev_3he-oszillationen,anderson_bec-interference,levy_bec-josephson-ac-bc,lagoudakis_polariton-josephson}, 
and even superfluid properties can be ascribed to most of these systems 
\cite{allen_4he-flow,kapitza_4he-viscosity,peshkov_tempcite,borovik_3he-induction,raman_bec-critical-velocity,amo_polariton-fluid}. 
Such hallmarks of a true BEC state are all based on the existence of 
a macroscopic phase-coherent state, but none of these has ever been 
observed in any insulating quantum spin system. 

In this article we propose an experiment to probe the macroscopic 
phase coherence in insulating spin systems by observing the a.c. 
Josephson effect across a weak link. A successful detection of this 
effect would represent a direct proof for the existence of a 
macroscopic quantum state. We start with 
a discussion about the macroscopic wavefunction and the chemical 
potential of a triplon BEC in a magnetic insulator at zero temperature 
(sections~\ref{Wavefunction} and ~\ref{chemPot}). We then consider 
the Josephson equations for a system of two weakly coupled triplon 
BEC's (section~\ref{Josephson}) and show how the a.c. Josephson effect 
can be identified in principle. In section~\ref{constraints} we discuss several contraints for a system with
realistic material parameters, and we finally conclude under realistic 
assumptions that such an experiment is indeed feasible.

\section{The macroscopic wavefunction of a triplon BEC}
\label{Wavefunction}

The amplitude of the macroscopic wave function 
\(\psi=\sqrt{n}e^{i\phi}\) describing a Bose-Einstein condensate of 
\(N\) triplons in a magnetic insulator at zero temperature is related to the density 
\(n\), here defined as \(n = N/N_d\) (with \(N_d\) the total number of dimers)
so that \(0 \leq n \leq 1\). To first approximation 
and near the critical field \(H_c\), this density is proportional to the 
longitudinal magnetic moment \(M_z = 
g\mu_BnN_d = g \mu_BN\) \cite{nikuni_tlcucl3_magnetization,matsumoto_tlcucl3_pressure},
where \(g\) is the Land\'e \(g\)-factor and \(\mu_B\) the Bohr 
magneton. As the equilibrium \(M_z\) is a well-defined quantity for a 
fixed value of the external magnetic field \(H\) in the thermodynamic 
limit, 
the triplon number \(N=M_z/g\mu_B\) can be 
considered to be conserved. 

The phase \(\phi\) of the macroscopic wave function in 
magnetic insulators is closely related to (but not identical with) 
the angle \(\varphi\) of the transverse 
magnetic moments within the plane perpendicular to the main magnetic 
field \(H\) 
\cite{giamarchi_magneticisolator-bec,sonin_spincurrent,nogueira_spin_josephson_ferromag}. 
A plausibility argument for this fact is that the particle number 
\(N\) is canonically conjugate to the phase \(\phi\) of the 
macroscopic wave function on the one hand, but also (via the 
proportionality of \(N\) to the longitudinal magnetic moment \(M_z\) 
and therefore to the associated component of the angular momentum 
\(L_z\) parallel to \(H\)) to the angle variable \(\varphi\) in the 
plane perpendicular to \(H\) on the other hand. 
In contrast to the time dependent global phase \(\phi(t)\) of the 
macroscopic wavefunction, 
the angle \(\varphi\) between the transverse magnetic moments and the 
frame of reference given by the crystal lattice must be constant in time, because the 
Landau-Lifshitz-Gilbert equation 
does not allow for a perpetual spin precession of the staggered 
magnetization in 
the magnetic ground state. This apparent discrepancy can be easily 
resolved if one interprets the global phase of the macroscopic wave function as 
\(\phi(t)=\varphi-\omega t\) (with \(\omega\) provided by the 
time-dependent Schr\"odinger equation), 
i.e., the in-plane angle \(\varphi\) of the staggered magnetization 
corresponds 
to an undetermined phase constant \(\phi_{0}\). This interpretation 
of \(\phi\) and \(\varphi\) ensures 
that the expectation values of \(L_z\) and of the energy \(E\) per 
triplon are \(\langle{L_z}\rangle_{\psi}=\hbar\) and 
\(\langle{E}\rangle_{\psi}=\hbar\omega\), respectively. 

In an axially symmetric spin system, \(\varphi\) must be able to take 
any arbitrary constant value 
between 0 and \(2\pi \) \cite{sonin_spincurrent}, which is in 
certain contrast to experimental observations where \(\varphi\) seems 
to be locked to a material-specific value 
\cite{tanaka_tlcucl3,kofu_ba3cr2o8}. We will 
address this issue of a violated axial symmetry later on in section~\ref{Viol}. 
In the following we shall focus on the time evolution of the global phase \(\phi(t)\) of the 
macroscopic wave function because its relation to the angle \(\varphi\) does not play any further important role here.

\section{The chemical potential \(\mu\)}
\label{chemPot}

A decisive prerequisite for an a.c. Josephson effect to take place 
between two BEC is the presence of a steady non-zero difference 
\(\Delta\mu\) 
between the respective chemical potentials. In a
condensate of magnetic quasiparticles in an insulating spin system, 
this chemical potential is usually taken as \(\mu = 
g\mu_B\mu_0(H-H_c)\) 
\cite{matsubara,batyev,giamarchi_ladders,nikuni_tlcucl3_magnetization}. 
As in all BEC 
of weakly interacting light bosons in the dilute limit \(n\ll 1\), a 
non-zero 
\(\mu\) is related to \(n\) according to \(n = \mu/v_0\). The 
interaction constant \(v_0\) describes the repulsive hard-core 
interaction between the bosonic quasiparticles, and is in magnetic 
insulators determined by 
the finite inter-dimer interactions which result in the formation of 
dispersive energy bands of the triplet excitations 
\cite{giamarchi_ladders}. It is important to emphasize that the 
appearance of a non-zero value of \(\mu = dE/dN\) (with \(E\) the 
total energy of the condensate) is common to all interacting BEC that 
are 
treated within the Gross-Pitaevskii formalism 
\cite{pitaevskii_bec,ozeri_bec},
and is not restricted to the magnetic systems under study here.

\begin{figure}
\centering
 \includegraphics[width=0.5\textwidth]{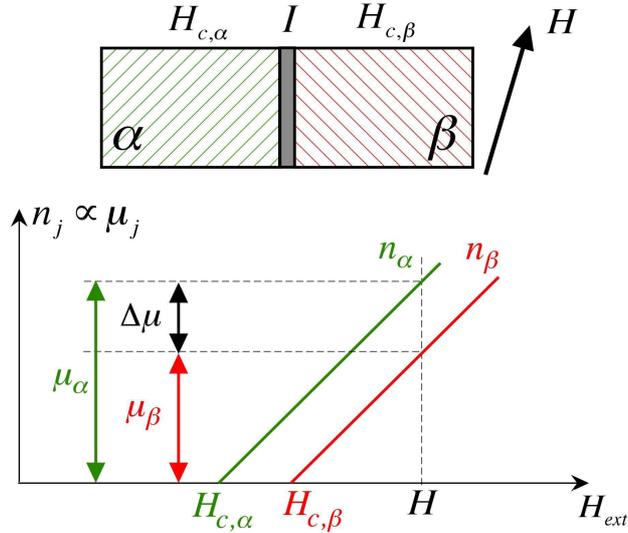}
 \caption{Sketch of an experiment in which two 
magnetic insulators with different critical fields \(H_{c,j}\) 
(\(j=\alpha,\beta\)) are weakly coupled to one another through a 
magnetically inert spacer layer \(I\) and are placed into an external 
magnetic field \(H>H_ {c,j}\). The respective condensate densities 
\(n_j\) and the chemical potentials \(\mu_j\) vary 
approximately linearly with \(H\).}
\label{fig1}
\end{figure}

In the following we aim to bring two of such condensates into contact 
in order to investigate possible analogues to the d.c. and a.c. 
Josephson effects. Therefore it is indispensable to examine 
whether or not the above definition of the chemical potential \(\mu\) 
is actually applicable to treat such an experiment correctly. If we 
consider, for example, two different magnetic insulators (\(\alpha\) 
and \(\beta\)) with different critical magnetic fields 
\(H_{c,\alpha}\) and \(H_{c,\beta}\) that are placed into a common 
external magnetic field \(H\), an apparently constant difference 
\(\Delta\mu = \mu_\alpha-\mu_\beta = 
g\mu_0\mu_B(H_{c,\beta}-H_{c,\alpha})\) is maintained by the external 
magnetic field (see Fig. \ref{fig1}). This situation is formally 
equivalent to a device composed of two pieces of the same material 
with a single \(H_c\), but placed into two different magnetic fields 
differing by \(\Delta H\). As we shall see later on, the necessary 
field 
gradient for the latter type of experiment to be successful is of 
the order of \(\mu_0\Delta H 
\approx 0.5\,\text{T}\) along a length of \(\approx 0.5\,\text{nm}\). 
This is technically out of reach, and we therefore do not consider 
this scenario any further.

This non-zero difference \(\Delta\mu\) is not simply a result of 
choosing different reference points of the energy scale for the two 
condensates. Firstly, the difference in the critical fields 
\(H_{c,j}\) (\(j = \alpha\text{ or }\beta\)) originally stems from a 
difference in microscopic inter- and intra-dimer coupling constants 
that determine the individual energy gaps, i.e. the energy separation 
of the respective ground-state singlet and the triplet states in zero 
magnetic field. To close these energy gaps, different external 
magnetic fields \(H_{c,j}\) must be applied beyond which the 
respective ground-state triplons condense. On an absolute 
energy scale (with the zero point chosen for a state with isolated 
spin-\(1/2\) particles in vacuum and in zero magnetic field) the 
condensation occurs at the 
energy level of the respective singlet states, i.e., at \(E_j = -3/4 
J_j\) per dimer, where \(J_j\) denote the respective intra-dimer 
coupling energies. Secondly and most importantly, the relevant 
quantity that enters the problem of treating the Josephson effect is 
the relative energy gain (loss) \(dE/dN\) upon creation (or 
annihilation) of 
one triplon quasiparticle to (or from) the condensate, from (or into) 
the singlet sea, which is actually \(n_jv_{0,j} = \mu_j = g_j 
\mu_B\mu_0(H-H_{c,j})\) per dimer\footnote{This result 
for \(\mu\)  can be illustrated by identifying \(E\) with the 
energy of the magnetized system in a field \(H\), \(E = 
\mu_0\int_{H_c}^H M_{z}(H')\,\mathrm 
{d}H'\) with \(M_{z}=g\mu_BN\) and 
\(N=g\mu_B\mu_0\left(H-H_{c}\right)N_{d}/v_{0}\), so that 
\(E(N)=N^2v_{0}/2N_{d}\) and \(\mu=\mathrm{d}E/\mathrm{d}N=nv_{0}\).}
\cite{affleck_antiferromag_bose,giamarchi_ladders,nikuni_tlcucl3_magnetization,giamarchi_magneticisolator-bec}. 
Therefore, the \(\mu_j\) (and along with them the condensate 
densities \(n_j = \mu_j/v_{0,j}\)) are indeed different in a common 
external magnetic field that exceeds both \(H_{c,\alpha}\) and \(H_{c,\beta}\) 
(see Fig. \ref{fig1}), and \(\Delta\mu\) can then be considered 
as the analogue to an external voltage controlling the difference 
between the chemical potentials in a superconducting Josephson 
junction.

\section{Josephson effects}
\label{Josephson}
\subsection{The Josephson equations}

We now apply this concept to a system of two dimerized spin systems 
at zero temperature and with different 
critical fields \(H_{c,\beta}   > H_{c,\alpha}\)  beyond which the 
respective magnetic quasiparticles are supposed to condense. We 
initially assume a perfect axial symmetry of the two magnetic systems 
with respect to the direction of the external magnetic field, 
but we will discuss the case of violated axial symmetry later on in section~\ref{Viol}. The 
boundary layers of the two systems are assumed to be weakly coupled 
to one another, and we place the device into a magnetic field \(H 
> H_{c,j}\) (see Fig. \ref{fig1}). For simplicity, we also 
assume \(v_{0,\alpha} \approx v_{0,\beta} = v_0\) and \(g_\alpha 
\approx g_\beta = g\), which is not essential for the main conclusion 
of our consideration, however.

In an approach introduced by Feynman \cite{feynman-lectures} to 
explain the Josephson effects across a weak link between two 
superconductors one considers the macroscopic wave functions 
\(\psi_j=\sqrt{n_j}e^{i\phi_j}\) on opposite sides of the junction, 
and treats the weak coupling between 
them according to \(i\hbar\frac{\partial}{\partial 
t}\psi_{\alpha;\beta}=\mu_{\alpha;\beta}\psi_{\alpha;\beta}+K\psi_{\beta;\alpha}\), 
where \(K\ll\mu_j\) is a phenomenological coupling 
constant. The resulting differential equations for the number of 
particles 
\(n_j\) occupying the respective macroscopic quantum states and for 
the corresponding phase difference 
\(\Delta\phi=\phi_\beta-\phi_\alpha\) become

\begin{subequations}
\begin{align}
&\frac{\partial n_{\alpha}}{\partial t}=-\frac{\partial n_{\beta}}{\partial 
t}=\frac{2K}{\hbar}\sqrt{n_{\alpha}n_{\beta}}\sin\left(\Delta\phi\right),\label{eqn:dichtevar}\\
&\frac{\partial\Delta\phi}{\partial t}=\frac{\mu_{\alpha}-\mu_{\beta}}{\hbar}-\frac{K}{\hbar}\frac{n_{\alpha}-n_{\beta}}{\sqrt{n_{\alpha}n_{\beta}}}\cos\left(\Delta\phi\right).\label{eqn:phasenvar}
\end{align}
\end{subequations}

\noindent For \(H_{c,\beta}>H_{c,\alpha}\), the solution 
of Eq. \ref{eqn:phasenvar} in the weak-coupling limit 
\(K\ll\mu_{\beta}\) is, to first approximation, 
\(\Delta\phi\approx\Delta\phi_0+\omega_\text{a.c.}t\) with 
\(\omega_\text{a.c.}\approx\Delta\mu/\hbar\) and an undetermined 
constant phase difference \(\Delta\phi_0\). The resulting variations 
in \(n_j(t)\) are very small, i.e, by a factor \(K/\Delta\mu\) 
smaller than their time-averaged values.

\subsection{The d.c. and a.c. Josephson effects}

\begin{figure}
\centering
 \includegraphics[width=0.5\textwidth]{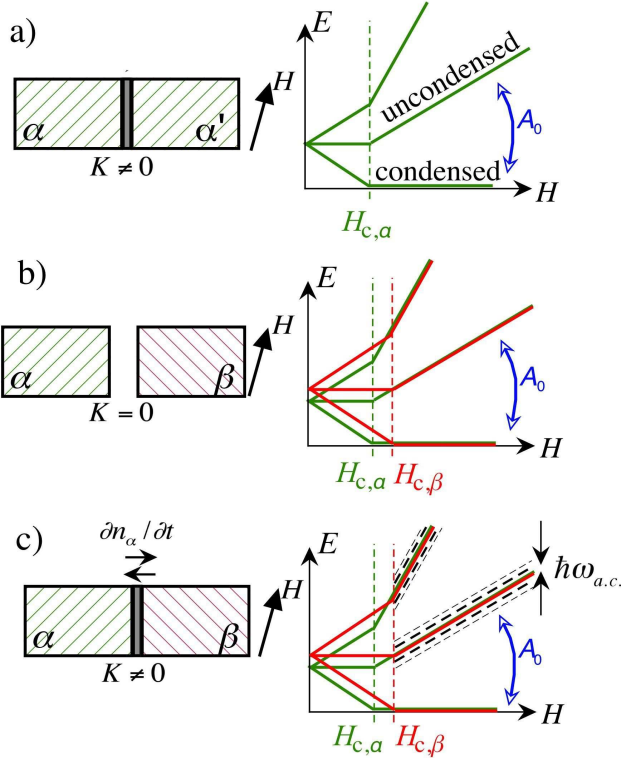}
 \caption{Sketch of the expected energy scheme (right 
panels) for a) two identical weakly coupled magnetic insulators, b) two uncoupled magnetic insulators with 
different critical fields, and c) two weakly coupled magnetic 
insulators with different critical fields (a.c. Josephson effect), 
all for \(H > H_{c,j}\) and measured relative to the respective 
ground-state energies. In the case c), the ESR-active mode \(A_0\) 
should split by a detectable amount \(\hbar\omega_\text{a.c}\) (see 
text). Arrows in the left 
panels symbolize the quasiparticle currents related to \(\partial 
n_\alpha/\partial t\) that can be ascribed to the spacer region (see 
text).}
\label{fig2}
\end{figure}

In the case of identical systems (\(\Delta\mu=0\) and 
\(n_\alpha=n_\beta\), see Fig. \ref{fig2}a) a constant 
\(\Delta\phi(t)=\Delta\phi_0=\phi_{0,\beta}-\phi_{0,\alpha}\)
would represent a constant quasiparticle current \(\partial 
n_\alpha/\partial t\) across the weak link (d.c. Josephson effect). 
As the constants \(\phi_{0,j}\) correspond to the in-plane angles 
\(\varphi_j\) of the respective transverse magnetic moments that will 
tend to align in the coupled device, we expect \(\Delta\phi_{0}=0\) and 
therefore \(\partial n_j/\partial t = 0\).

In a device with two magnetic insulators with different critical 
fields, however, \(\Delta\mu\neq0\) and therefore 
\(\omega_\text{a.c.}\neq0\) (Fig. \ref{fig2}c). As one would expect it from the analogy 
with superconductors and superfluids, the resulting variation 
\(\partial n_\alpha/\partial t\)  given by Eq. \ref{eqn:dichtevar} 
represents an oscillation in time with the field-independent leading 
frequency \(\omega_\text{a.c.}\)  (a.c. Josephson effect).

The analogous situation in a superconducting Josephson device leads 
to the appearance of Josephson electrical currents. If connected to 
an external charge reservoir and fixing \(\Delta\mu\) to a constant 
value, the net charge-carrier density remains constant despite a 
nonzero \(\partial n_\alpha/\partial t\), but a measurable electrical 
current proportional to \(\partial n_\alpha/\partial t\) flows from 
the charge reservoir and the attached leads through the whole 
device \cite{feynman-lectures}.

With a junction composed of insulating spin systems it is 
the external magnetic field that maintains the difference between the 
chemical potentials to a certain value. This field fully penetrates 
the whole sample volume, and it therefore entirely 
"short-circuits" each of the two spin systems separately, thereby 
keeping the respective total numbers of quasiparticles \({n_j}\)  
constant. As a consequence, there is no directional macroscopic 
quasiparticle current flowing \textit{within} the two individual 
branches on both 
sides of the device. One can think of the magnetic field as a 
quasiparticle 
source (or sink) that replaces (or removes) those quasiparticles that 
are crossing the weak link. It is only by virtue of the phase 
difference \(\Delta\phi\) between the two materials that a quasiparticle current can 
still be formally ascribed to the magnetically inert spacer region 
(see Fig. \ref{fig2}c), but it may be 
impossible to detect it directly in an experiment.

\subsection{Experimental manifestation of the a.c. Josephson effect}

To second-order approximation (\(K\neq0\) with \(H_{c,\alpha}\neq 
H_{c,\beta}\) and therefore \(n_\alpha\neq n_\beta\)), Eq. 
\ref{eqn:phasenvar} denotes a narrowband frequency 
modulation of the phases 

\begin{equation}
\phi_{\alpha;\beta}(t)\approx\phi_{0,\alpha;\beta}-\frac{\mu_{\alpha;\beta}}{\hbar}t-\frac{K}{\Delta\mu}\sqrt{\frac{n_{\beta;\alpha}}{n_{\alpha;\beta}}}\sin(\omega_\text{a.c.}t+\Delta\phi_0),\label{eqn:phi}
\end{equation}

\noindent with a 
modulation frequency 
 \(\omega_\text{a.c.}\) and a modulation index of the order of 
 \(K/\Delta\mu\). This modulation of \(\phi_{j}(t)\) has profound 
 consequences on the macroscopic wave functions 
 \(\psi_j=\sqrt{n_j}e^{i\phi_j}\) describing the two condensates, 
because equally spaced sidebands should appear in the 
energy spectrum of the coupled device. This dynamic effect should 
manifest itself in a corresponding splitting of all energies that are 
associated with transitions from the (condensed) ground state to 
(uncondensed) excited states (see Fig. \ref{fig2}c). Such a 
splitting is absent for two uncoupled 
magnetic insulators with different critical fields (Fig. 
\ref{fig2}b) and, of course, also in a scenario where no 
macroscopic phase coherence is present at all. Therefore a successful 
experiment in a coupled system that can probe the occurrence of 
sidebands with separation \(\hbar\omega_\text{a.c.}\) would 
represent a very strong experimental support for the existence of a 
state with 
macroscopic phase coherence. 

To be more specific, a high-resolution 
electron-spin resonance (ESR) measurement of the transitions between 
the ground state considered here, and the nearest 
excited triplet states (referred to as \(A_0\) in Ref. 
\cite{kimura_tlcucl3_esr} and \(E_0(Q)\) in Ref. 
\cite{matsumoto_spindimer_esr}), or a one-magnon Raman 
experiment on the so-called \(E_-(Q)\)-mode 
\cite{kuroe_tlcucl3_raman}, should reveal a characteristic splitting 
of the corresponding modes by \(\hbar\omega_\text{a.c.}\) which is 
related to the a.c. Josephson effect (see Fig. \ref{fig2}c). This 
is a 
central result of this work, 
and we shall discuss in the following whether or not the proposed 
a.c. Josephson effect can be observed in a device with realistic 
material parameters.

\section{Constraints in real systems}
\label{constraints}
\subsection{The lifetime of phase coherence}

We may expect a successful experiment only if the separation 
\(\hbar\omega_\text{a.c.}\) between associated
sidebands in the energy spectrum is comparable to or larger than 
their linewidth which can be related to the inverse lifetime 
\(\tau_{pc}^{-1}\) of phase coherence and to the inverse lifetime 
\(\tau_{qp}^{-1}\) of the magnetic quasiparticles involved, i.e., 
\(\omega_\text{a.c.} \gtrsim max(\tau_{pc}^{-1},\tau_{qp}^{-1})\). 
The lifetime 
\(\tau_{pc}\) has, to the best of our knowledge, 
not yet been estimated for triplon BEC's, but we may initially 
compare it to \(\tau_{qp}\), i.e., that of the \(S = 1\) 
quasiparticles. Scattering processes on magnons, phonons or 
impurities can be important limiting factors, but there is no 
fundamental principle 
that would restrict it to inaccessibly short time scales at low 
enough 
temperatures. Corresponding values for the quasiparticle lifetime 
\(\tau_{qp}\) 
ranging from \(>10^{-11}\,\text{s}\) up to 
\(\approx5\times10^{-11}\,\text{s}\) 
can be inferred from precise inelastic neutron-scattering 
measurements of the magnon linewidth in the spin-dimerized compounds 
TlCuCl\(_3\) \cite{ruegg_dimer-systems} and 
(C\(_4\)H\(_12\)N\(_2\))Cu\(_2\)Cl\(_6\) 
\cite{stone_2dantiferromagnet}, respectively, taken at low 
temperatures for quasiparticles that are not part of the condensate, 
and from the ESR linewidth of the \(A_0\) mode observed in 
TlCuCl\(_3\) 
\cite{kimura_tlcucl3_esr}. 

It is conceivable that the actual lifetime \(\tau_{pc}\)
of phase coherence in the BEC is even 
considerably longer than 
that of its constituting particles \(\tau_{qp}\), e.g., thanks to the 
formation of energetic 
barriers of topologic origin that may lead to a strong suppression of 
dissipation, i.e., to the occurrence of true spin superfluidity 
\cite{sonin_spincurrent}. 
It is worth mentioning here that the a.c. Josephson oscillations that 
have been seen in an experiment on an exciton-polariton 
Josephson 
junction were clearly observable although the oscillation frequency 
(\(\omega_\text{a.c.}\approx 4\times10^{11}\,\text{s}^{-1}\)) was 
very near the inverse lifetime of the respective bosonic 
quasiparticles 
(\(\tau_{qp}^{-1}\approx 3\times10^{11}\,\text{s}^{-1}\)) 
\cite{lagoudakis_polariton-josephson}, 
but still much larger than the inverse 
of the surprisingly long measured phase-coherence time 
(\(\tau_{pc}^{-1} \approx 7\times10^{9}\,\text{s}^{-1}\) 
\cite{love_polariton_decoherence}, i.e., 
\(\tau_{pc} >> \tau_{qp}\)). 
To be on the safe side we conservatively choose 
\(\,\omega_\text{a.c.}\)\(\approx 10^{11}\,\text{s}^{-1} > 
\tau_{qp}^{-1}\) for our 
proposed experiment, which 
corresponds to \(\mu_0(H_{c,\beta}-H_{c,\alpha}) \approx 
0.5\,\text{T}\).

\subsection{Violated axial symmetry}
\label{Viol}

So far we have considered only perfectly axially symmetric materials. 
Any violation of axial symmetry of the magnetic exchange 
interaction leads to the occurrence of an anisotropy gap 
\(\it{\Delta}\) in the energy spectrum \(E(k)\) 
\cite{sirker_tlcucl3}, thereby lifting the Goldstone linearity for 
\(k\rightarrow0\), fixing the phases \(\phi_{0,j}\) to constant 
values \cite{affleck_antiferromag_bose,dellamore_tlcucl3-symmetry-breaking} 
and leading to well-defined magnetic structures with in-plane 
angles \(\varphi_j\) that are locked to the crystal lattice 
\cite{tanaka_tlcucl3,kofu_ba3cr2o8}.
This gap has been estimated to 
\(\it{\Delta}(H)=\sqrt{8\tilde{\gamma}\mu(H)}\), where 
\(\tilde{\gamma}> 0\) is a measure for the exchange-interaction 
anisotropy in the plane perpendicular to \(H\) \cite{sirker_tlcucl3}.
For experimental energies larger than \(\it{\Delta}\), however, a 
quasi-linear Goldstone mode is recovered \cite{ruegg_tlcucl3_nature}, 
the effects of axial anisotropy are expected to be smeared out on 
short enough time scales, and we can ascribe an upper 
limit \(\tau_{\it{\Delta}}=\hbar/\it{\Delta}\) to the lifetime of 
phase coherence \cite{dellamore_tlcucl3-symmetry-breaking} so that we 
must additionally require \(\omega_\text{a.c.} > 
\tau_{\it{\Delta}}^{-1}\). 

\subsection{The weak link}

In a real experiment, 
\(\Delta\mu=g\mu_0\mu_B(H_{c,\beta}-H_{c,\alpha})\) and the coupling 
constant \(K\) are determined by the junction properties. The 
dimensions of the weak link should not significantly exceed the 
healing length of the condensate (the analogue to the coherence 
length in 
superconductors). For weakly interacting Bose gases, this healing 
length is \(\xi=\hbar/\sqrt{2nv_0m^*}\) (where \(m*\) is the 
effective mass of the quasiparticles) \cite{bogoliubov_tempcite} and 
it diverges as the temperature is increased towards the Bose-Einstein 
condensation temperature where \(n\) vanishes.

The simplest version of such a weak link may consist of a slit that 
is 
narrow enough to allow for a small but finite magnetic coupling and 
therefore for a tunnelling of magnetic quasiparticles. Such a 
junction might also be manufactured by the layered growth of a 
magnetic insulator (\(\alpha\), with \(H_{c,\alpha}\)) containing a 
certain species of magnetic ions forming spin-1/2 dimers, followed by 
one or a few layers of an isostructural compound with nonmagnetic 
ions, and by further adjacent layers of another isostructural 
magnetic insulator (\(\beta\), with \(H_{c,\beta}\neq 
H_{c,\alpha}\)).

\subsection{A realistic toy model: (Ba,Sr)\(_3\)Cr\(_2\)O\(_8\)}

To obtain numeric values, we consider a model device composed of 
Ba\(_3\)Cr\(_2\)O\(_8\) (\(\alpha\)) with 
\(\mu_0H_{c,\alpha}\approx12\,\text{T}\) \cite{kofu_ba3cr2o8}, 
separated by nonmagnetic isostructural Ba\(_3\)V\(_2\)O\(_8\) 
\cite{suesse_citetemp} from a compound 
Ba\(_{3-x}\)Sr\(_x\)Cr\(_2\)O\(_8\) (\(\beta\)) 
\cite{aczel_sr3cr2o8}\footnote{We have already verified the existence 
of the solid-solution series Ba\(_{3-x}\)Sr\(_x\)Cr\(_2\)O\(_8\) for 
polycrystalline samples, but we have no further information about 
their magnetic properties.} with \(x\) chosen in such a way to 
achieve a larger \(\mu_0H_{c,\beta}\), but still much smaller than 
\(\mu_0H_ {c2} = 30\,\text{T}\) of fully 
stoichiometric Sr\(_3\)Cr\(_2\)O\(_8\) with \(x 
= 1\) \cite{aczel_sr3cr2o8}. If we achieve a \(\mu_0H_{c,\beta} 
\approx 12.5\,\text{T}\), 
the expected characteristic frequency becomes 
\(\omega_\text{a.c.}\approx 10^{11}\,\text{s}^{-1}\). With an 
external magnetic field \(\mu_0H \approx 13\,\text{T}\), and taking 
\(v_0/k_B \approx 8.7\,\text{K}\) with \(g \approx 2\) 
\cite{dodds_ba3cr3o8} we 
have \(n_\alpha \approx 0.15\) and \(n_\beta 
\approx 0.08\). The choice of a small \(H-H_{c,j}\) (i.e., 
\(\mu_j\ll v_0\)) is primarily necessary to fulfill the dilute-limit 
condition \(n_j\ll 1\), but it may not be crucial for observing the 
essence of the predicted effect. With \(m^*\approx 1.5\times10^{-27}\,\text{kg}\) and 
a mean distance \(d\approx 0.6\,\text{nm}\) between the dimers 
\cite{kofu_ba3cr2o8,dodds_ba3cr3o8}, the resulting values for the 
healing lengths at \(T = 0\) are small 
(\(\xi_\alpha\approx0.4\,\text{nm}\) and 
\(\xi_\beta\approx0.6\,\text{nm}\approx d\)), but a very narrow slit 
or a weak link composed of one or only a few unit cells of the 
nonmagnetic compound could be adequate. If still larger values of 
\(\xi_j\) are required, \(n_j\) can be diminished by further reducing 
\(\Delta\mu\) and \(\mu_\beta\). Alternatively, the experiment can be 
performed near the condensation temperature as it has been done in 
superfluid \(^4\)He to make the a.c. Josephson effect observable 
\cite{sukhatme_he-josephson}, possibly at the cost of reducing the 
lifetime \(\tau_{qp}\) of the triplon quasiparticles. 

With the 
material parameters used for the experiment described above and 
\(\,\omega_\text{a.c.}\)\(\approx 10^{11}\,\text{s}^{-1}\), 
\(\tilde{\gamma}\) should be below \(\approx 4\,\mu\text{eV}\). The 
exchange-interaction anisotropy has been estimated to 
\(\tilde{\gamma} \approx 16-30\,\mu\text{eV}\) in TlCuCl\(_3\) 
\cite{dellamore_tlcucl3-symmetry-breaking,kolezhuk_anisotropie-dimerdynamik}, 
but to only \(\tilde{\gamma} \approx 1\,\mu\text{eV} < 
4\,\mu\text{eV}\) in BaCuSi\(_2\)O\(_6\) 
\cite{sebastian_anisotropie-bacusi2o6} and particularly in 
Ba\(_3\)Cr\(_2\)O\(_8\) \cite{kofu_ba3cr2o8}, so that the proposed device 
would fulfil all of the requirements listed above.

\section{Concluding remarks}
 
To conclude these numeric estimates we want to state that other 
material properties, such as the presence of a strong 
Dzyaloshinski-Moriya interaction that we have not considered here, 
may still inhibit the formation of a phase coherent condensate 
\cite{sirker_tlcucl3} and therefore make the observation of the a.c. 
Josephson effect impossible. However, we classify such factors as not 
intrinsic to the problem, and a proper choice of compounds with 
suitable material parameters should make the generation and the 
observation of the a.c. Josephson effect in magnetic insulators 
feasible in principle. Although we have chosen dimerized spin systems 
operating near their lower critical fields as a model for the present 
consideration, all the above arguments should also hold near the 
respective saturation fields as a consequence of the particle-hole 
symmetry of the problem \cite{giamarchi_ladders}, and for all other 
types of insulating spin systems that are supposed to show a 
field-induced BEC of magnetic quasiparticles. It is conceivable that 
Josephson-like phenomena can even naturally occur in certain quantum 
magnets, e.g., in a system showing 
an intrinsic modulation of the boson density (and therefore of the chemical 
potential) due to inequivalent planes hosting the triplon condensate. 
Such a situation has been reported for BaCuSi\(_2\)O\(_8\), 
in which the average boson density strongly varies along the \(c\)-axis 
\cite{Kraemer,Mila}.  

We finally mention here an interesting analogy between the case of an 
ESR measurement on weakly 
coupled magnetic insulators as discussed above, and corresponding 
microwave stimulated
experiments on ferromagnetic 
films which are separated by normal-metal spacers 
\cite{heinrich_dynamicexchangecoupling}. 
In both cases, there exists a dynamic coupling between two different 
materials through a magnetically inert spacer layer, 
thereby 
altering the dynamics in both systems as compared to the 
uncoupled situation.

In summary, we suggest that two weakly linked magnetic insulators 
with different critical fields \(H_{c,j}\) that are placed into a 
suitably chosen 
external magnetic field should show an altered energy spectrum as 
compared to the uncoupled limit, with additional sidebands 
that are separated by \(\hbar\omega_\text{a.c.}=
g\mu_B\mu_0(H_{c,\beta}-H_{c,\alpha})\). 
These sidebands are a manifestation 
of the a.c. Josephson effect and can be tested, for example, in a high-resolution 
ESR experiment. 
Assuming realistic material parameters and considering several 
constraints, we conclude that this effect can indeed take place in a 
real system. A corresponding successful experiment would represent 
a very strong experimental support for the existence of a state with 
macroscopic phase coherence.

\section{Acknowlegdements}
We thank to A. Rosch, J. Roos, T. Giamarchi, M. Matsumoto and A. 
Rakhimov for stimulating discussions. This work was supported by the 
Swiss National Foundation Grant. No. 21-126411.

\end{document}